\documentstyle[epsf,fleqn,aps]{revtex}  

\def\Section#1{}
 
\def\beq{\begin{equation}}
\def\eeq{\end{equation}}
\def\bea{\begin{eqnarray}}
\def\eea{\end{eqnarray}}

\begin{document}
\twocolumn[\hsize\textwidth\columnwidth\hsize\csname 
@twocolumnfalse\endcsname
\title{Massive and Massless Behavior in Dimerized Spin Ladders}
 
\author{D.C.\ Cabra$^{1,\#}$ and M.D.\ Grynberg$^2$}

\address{$^1$Physikalisches Institut der Universit\"at Bonn, Nussallee 
12, 53115 Bonn, Germany\\
$^2$Departamento de F\'{\i}sica, Universidad Nacional de la Plata,
      C.C.   67, (1900) La Plata, Argentina}

\maketitle
\date{\today}
\maketitle

\begin{abstract}
We investigate the conditions under which a gap vanishes in the spectrum
of dimerized coupled spin-1/2 chains by means of Abelian bosonization
and Lanczos diagonalization techniques.
Although both interchain ($J'$) and dimerization ($\delta$) couplings favor
a gapful phase, it is shown that a suitable choice of these interactions 
yields massless spin excitations. We also discuss the influence of different
arrays of relative dimerization on the appearance of non-trivial 
magnetization plateaus.

\vspace{10 pt}
\hspace{0.5cm} Published in Phys. Rev. Lett. {\bf 82}, 1768 (1999).
\vspace{-12 pt}
\end{abstract}
\vskip2pc]

Current developments in the field of spin ladders have revealed
intriguing features of low-dimensional quantum antiferromagnets 
(AF) \cite{Elbio}. The appearance of plateaus in the magnetization 
curves of these systems has received much attention  from both 
the theoretical and experimental side \cite{CHP}. A necessary
quantization condition for the appearance of such plateaus in generic 
ladders was derived in \cite{CHP} (see also  \cite{OYA} for the case of 
1D chains and \cite{Totsuka,Totsuka2,CG,Andreas} for other particular cases).
In a very recent paper \cite{Shiramura},  an unexpected phenomenon has been 
observed in the magnetization curves of 
NH$_4$CuCl$_3$ at high magnetic fields.
The crystal structure of this material at room temperatures is known to be
composed of double chains (2-leg ladders), with three different nearest
neighbor interactions. Thus, from previous theoretical studies of
such arrange of couplings in ladders systems, plateaus at $\langle
M\rangle =0$ and $1/2$ are expected to be visible at sufficiently low
temperature. However, measurements performed in \cite{Shiramura} at very 
low temperatures, clearly show two net plateaus at 
$1/4$ and $3/4$, while {\it no} plateaus are observed at the theoretically 
expected values $\langle M \rangle =0$ and $1/2$.

As a first step to reconcile some of these facts with the current 
understanding of ladder materials, in this work we study two important 
issues namely, (i) the possibility of closing the gap in a two-leg dimerized 
ladder by a combined effect of the dimerization and the interchain coupling
and, (ii) the emergence of a plateau at $\langle M \rangle =1/2$ depending 
on the realization of the relative dimerization (see Figs.\ 1(a)-(b)).
Though (i) was studied on a qualitative level using non-linear sigma 
model techniques \cite{Sierra1}, here we present a quantitative and more 
systematic treatment, whereas (ii) is analyzed for the first time 
in this work. These two issues could shed light in the study 
of the experimental measurements such as those performed in 
\cite{Shiramura} as well as in related aspects of CuGeO chain compounds 
\cite{Boucher}, in which dimerization becomes staggered between weakly 
coupled chains.

As is well known, the isotropic spin-1/2 Heisenberg chain is already in a 
critical state. Thus any relevant perturbation, 
such as the Peierls dimerization 
instability \cite{Cross} or a weak interchain coupling 
\cite{DRS}, \cite{SNT}, can drastically alter the nature
of the ground state, whereas a massive spin gap excitation appears 
simultaneously in the energy spectrum. Interestingly, it was 
suggested that a combined
effect could lead to a massless regime \cite{Sierra1}.

In the present work we study this issue quantitatively by
means of both bosonization and  numerical techniques, and find that there
exists indeed a fine-tuning of the couplings responsible for the appearance
of massless lines. Though this is a fine-tuning effect, 
our result implies that there is a 
{\it finite} region of the coupling parameter 
space where the gap is expected to be small. Thus, when the latter
becomes comparable with thermal fluctuations, measurement attempts
of  zero magnetization plateaus could be smeared out even at
low temperatures.

In analyzing  the abelian bosonization of spin-$1/2$ Heisenberg ladders,
we follow a similar methodology developed as in \cite{CHP,Totsuka2}, 
\cite{SNT},\cite{LeH},\cite{SchBo}, which is particularly 
suitable to elucidate the behavior of weak coupling regimes.
Specifically, here we study two dimerized spin chains 
interacting through a Hamiltonian 
\beq 
\label{H12}
H = \sum_{a,n} \, J_n^{(a)} \; \vec
S_n^{\;(a)} \cdot \vec S_{n+1}^{\;(a)}\,+\,
J' \sum_n \, \vec S_n^{(1)} \cdot \vec S_n^{\,(2)}\,, 
\eeq
$(a =1,2)$, where the $\vec S_n$ denote spin-$1/2$ operators.  
For {\it staggered} ladders, e.g.\ Fig.\ 1(a), the array of coupling exchanges 
are set as $J_n^{(2)} \equiv J_{n+1}^{(1)}$, and parametrized by 
$J_n^{(1)}=J\left[\,1+(-1)^n \delta \,\right]$ say for chain (1),
whereas for the non-staggered situation we just set $J_n^{(2)} \equiv
J_n^{(1)}$, as shown in Fig.\ 1(b). To maintain pure AF and non-frustrated
exchanges, the dimerization parameter is kept bounded 
by $\vert \delta \vert < 1$ throughout the $2 L$ spins of the ladder 
length with periodic boundary conditions ($L$ even). 

On general grounds \cite{CHP}, it is 
expected that gapful magnetic excitations should appear for all 
magnetizations $\langle M \rangle \equiv \frac{1}{L}\, \sum_n  
\langle S_n^{z (1)} \!+ \!S_n^{z (2)} \rangle$, 
satisfying the quantization condition
$p N (1- \langle M \rangle)/2 \in {\cal Z},$
where $p$  stands for the polymerization \cite{CG} and $N$ for the number 
of chains coupled together.
This would imply the existence of plateaus at $\langle M \rangle = 0$
and, if dimerization occurs ($p=2$), also at $\langle M \rangle = 1/2$, 
for a $N=2$-leg ladder material such as the one studied in \cite{Shiramura}. 
However, the experimental magnetization curve shows no plateaus at these 
two values of the magnetization, but instead at $\langle M \rangle = 1/4$ 
and $\langle M \rangle = 3/4$ \cite{Shiramura}.
To explain this apparent breakdown, we are especially 
interested to include  an homogeneous field term 
of the type $ - \,\frac{h}{2}\,\sum_n \, 
\left[ S_n^{z (1)} + S_n^{z (2)} \right]\,$,
so as to unravel the interplay between the above  kinds of coupling arrays 
and applied  magnetic fields $h$, say along the $z$-direction.

It is well known that the low-energy properties  of the Heisenberg chain,
($\delta = 0$), are described by a $c=1$  conformal field theory of a free
bosonic field compactified at radius $R=R(\langle M \rangle, \Delta)$ 
\cite{CHP,Totsuka,Totsuka2,LeH,SchBo},
for any given magnetization $\langle M \rangle$ and $XXZ$ anisotropy 
$\vert \Delta \vert < 1$ (see e.g.\  \cite{HaldXXZ}\,). 
The functional dependence of $R$ 
can be obtained using the exact Bethe Ansatz solution by solving a set of 
integral equations obtained in \cite{Woyna,Bogo} and \cite{bookK} 
(for a fuller review consult for instance Ref. \cite{CHP}\,).
The bosonized expression of the low-energy effective Hamiltonian for a single 
homogeneous chain,  ($\delta =0$), in the presence of an 
external magnetic field $h$ and with an $XXZ$ 
anisotropy $-1< \Delta < 1$,  can be shown to adopt the form
\beq
\bar{H}= \int {\rm d}x {\pi \over 2} \left\{ \Pi^2(x)/(4R)^2 +
   R^2 \left(\partial_x \phi(x)\right)^2
\right\}\,,
\label{2}
\eeq
with $\Pi =  \partial_x \tilde{\phi}$, and
$\phi = \phi_L + \phi_R$, $\tilde{\phi} = \phi_L - \phi_R$.
Here, the effect of the magnetic field $h$ and the $XXZ$ anisotropy  
$\Delta$ enter through the radius of compactification
$R$. This radius governs the conformal
dimensions, in particular the conformal dimension
of a vertex operator ${\rm e}^{i \beta \phi}$ is given by   
$\left({\beta \over 4 \pi R}\right)^2$.

In the limit of  both weak dimerization $\vert \delta \vert \ll 1$, 
and interchain coupling $J'/J\ll 1$,  the bosonized action reads
\bea
H_{int}\approx  \lambda_1
\sum_x \partial_x\phi^{(1)}\partial_x\phi^{(2)}+
\nonumber\\
\lambda_2 \sum_x
\cos(4k_F x+\sqrt{4 \pi}(\phi^{(1)}+\phi^{(2)}))+
\nonumber\\
\lambda_3 \sum_x 
\cos(\sqrt{4 \pi}(\phi^{(1)}-\phi^{(2)}))+
\nonumber\\
\lambda_4\sum_x
\cos(\sqrt{\pi}(\tilde\phi^{(1)}-\tilde\phi^{(2)}))+
\nonumber\\ 
\sum_{(a)=1}^2 \sum_x  \alpha^{(a)}(-1)^x \cos(2k_F (x+1/2)+
\sqrt{4 \pi}\,\phi^{(a)}(x))\,.
\label{Hint}
\eea
where $\lambda_i\propto J'/J$, $\alpha^{(1)} \propto \delta$
and $\alpha^{(2)} \propto \pm \delta$, (the minus sign corresponding to 
Fig.\ 1a), and the Fermi 
momentum $k_F$ is related to the total magnetization $\langle M \rangle$ 
via $k_F=(1-\langle M \rangle)\pi/2$. In (\ref{Hint}) a marginal term has 
been neglected for the sake of simplicity since it does not change our 
results.

\vspace{.2cm}

\noindent {\it (i) Closing of the gap}

\vspace{.2cm}

At zero magnetization  (i.e.\ $k_F=\pi/2$), all
terms in (\ref{Hint}) are commensurate and relevant.
Due to the $\lambda_1$ term, we have to  first diagonalize
the derivative part of the Hamiltonian which is achieved by
introducing new variables $\phi^{\pm} = (\phi^1 \pm \phi^2)/\sqrt{2}$.
The associated compactification radii are given by
$R_{\pm} = R \sqrt{1 \pm J'/( 4 J \pi^2 R^2)}$.

For AF interchain coupling, ($J'> 0$), and in the new basis,
the $\lambda_4$ term is the most relevant and orders the 
$\tilde{\phi}^-$ field (the dimerization terms mix the $\phi^+$
and $\phi^-$ variables and will be treated as a perturbation). 
The $\alpha$ interaction in (\ref{Hint}) is then wiped out after 
integrating  
the massive $\phi^-$ field, since it always contains a contribution 
from $\phi^-$. Thus we are left with the $\phi^+ $ field 
with a relevant interaction given by $\lambda_2$. Hence, this field 
is in general massive so we can expect a plateau at zero magnetization.
The interesting point here is that the $\alpha$ interaction 
generates radiatively a
term that can cancel the $\lambda_2$ perturbation,
(which is the one responsible for the mass of the symmetric boson).
{}From this plain weak coupling analysis we see that 
this happens on the critical line given by
\beq
\label{ft}
\frac{J'}{J} \propto \delta^2\,.
\eeq
Therefore, whenever this cancellation occurs we have one
massless degree of freedom corresponding to the 
symmetric combination of the original free bosons $\phi^+$.

To enable an independent check of this result, we now turn to a 
numerical finite-size analysis of the original ladder Hamiltonian. 
We direct the reader's attention  to Fig.\ 2 where we display 
the energy gap necessary to create an excitation with total spin $S=1$
in the staggered coupling array [Fig.\ 1(a)]. The results were obtained from 
exact diagonalizations of finite ladder lengths, $4 \le L \le 12$,
via a recursion type Lanczos algorithm \cite{Lanczos}.
In studying the mass gap extrapolations towards their 
thermodynamic limits (dashed curve of Fig.\ 2),
we fitted the whole set of finite-size results using a variety of
standard procedures; namely  linear, logarithmic and van den 
Broeck-Schwartz type methodologies of convergence \cite{Gutt}. 
Basically, they yield analogous results with at least 2 significant digits,
though within the neighborhood of the massless region finite size
effects become noticeable. In particular, this latter variation is reflected
in the localization of the critical line contained within the error bars
of Fig.\ 3. Nevertheless, at least in the weak coupling $(\delta,J')$ regime
we were able to obtain a reasonable agreement with Eq. (\ref{ft}).
The progressive departure from the latter however, yields the right critical
parameters. As  $\vert \delta \vert  \to 1$, we clearly  recover 
the ``snake chain''  limit where criticality is achieved when  
$J' = 2J$  [see Fig.\ 1(a)]. This strong coupling regime yields 
in turn the expected linear behavior in $(\vert \delta \vert -1)\,$.
 
Preliminary estimations of the sound 
velocity \cite{Cardy} near the ground state are consistent with a conformal 
anomaly $c=1$ throughout the critical line, except for $\delta = J' =0$,  
where the system has two decoupled bosons with $c=2$. 
Moreover, a low temperature computation of the magnetic susceptibility 
revealed that the latter is enhanced significantly on approaching the 
massless line  (\ref{ft}), thus lending further support to our bosonization 
approach.

Also, this effect is presumed to occur 
on the ferromagnetic side, $J' < 0$, where a critical line can emerge 
with usual dimerization as well [Fig.\ 1(b)\,].
In fact, a similar analysis using the perturbative RG approach
has been given in the first reference of \cite{SNT} and, indeed, the 
effect is observed numerically \cite{CG2}.

\vspace{.2cm}

\noindent {\it (ii) Plateau at $\langle M \rangle = 1/2$}

\vspace{.2cm}

Let us now consider Eq.\,(\ref{Hint}) at non-zero magnetization 
$\langle M \rangle$.  As was referred to above, this can 
be readily accounted through
the radius of compactification $R(\langle M \rangle, \Delta)$ resulting
from the effect of the magnetic field $h$ and the Fermi
momentum $k_F=(1-\langle M \rangle )\pi/2$. Then, the $\lambda_2$ and
$\alpha$ interactions are incommensurate, and we have in principle a
massless mode corresponding to the $\phi^+$ field. However an interesting
phenomenon can be observed: a plateau in the magnetization curve is
expected to appear at $1/2$ of the saturation value due to the appearance
of the radiative correction 

\beq \gamma \sum_{x=1}^L (-1)^x
\cos(4k_F(x+1/2)+\sqrt{8\pi}\phi^+). 
\label{plateau} 
\eeq
This term is generated from a combined effect of the dimerization on each 
chain ($\alpha^{(a)}$ interactions in (\ref{Hint}))  and the interchain 
interaction terms $\langle M \rangle \left(\cos(2k_Fx+\sqrt{4\pi}\phi^{(1)})+
\cos(2k_Fx+\sqrt{4\pi}\phi^{(2)})\right)$.

However, the presence of this radiatively generated term is strongly
dependent of the {\it manner} in which the dimerization is realized. In the
case of normal dimerization, one finds that 
$\gamma \propto \delta J' \langle M \rangle$, while in the staggered case 
it vanishes exactly, $\gamma \equiv 0$, due to the relative minus sign 
between $\alpha^{(1)}$ and $\alpha^{(2)}$. This effect, predicted from our 
weak coupling analysis can be clearly observed by comparing Figs.\ 
4(a) and 4(b) (see below).

The operator (\ref{plateau}) is relevant 
for any anisotropy $\Delta > 0$, where a Kosterlitz-Thouless (KT) transition 
is expected to take place. (More precisely, this KT point will depend also 
on the interchain coupling $J'$ through $R_{\pm}$

One can also understand this difference in a sort of strong coupling
limit, when $\vert \delta \vert \rightarrow 1$. 
It can be readily checked that the
normal dimerization displays an $\langle M \rangle =1/2$ plateau $\forall
h \in (2J,\, 4J)$ with $J'=J$, while there is no plateau in the staggered
array. This effect is therefore robust in the sense that it is valid for
all values of the couplings and could then be a useful tool for
experimentalists \cite{Shiramura} to decide whether the former or the
latter exchange topology is actually realized in a given compound. 

To test the correctness of this scenario, finally we recur to the Lanczos
algorithm, to obtain the lowest  energy eigenvalues on each 
magnetization subspace $S^z =\{0,1,\,...\,,L\}$. 
Using ladders up to $L=12$, in Figs.\ 4(a) and (b) we
display respectively the ground state magnetization curves 
of the staggered and normal coupling arrays.
By setting $J'/J =2$ and $\delta =0.5$, our numerical 
analysis supports entirely the bosonization 
picture as {\it no} plateau shows up for staggered arrays at  
any finite magnetization [Fig.\ 4(a)]. In contrast, the 
normal array [Fig.\ 4(b)] clearly exhibits the expected 
$\langle M \rangle =1/2$ plateau. Furthermore, the {\it robustness} 
of this result  was checked numerically for a variety of dimerization 
($\delta$) and interchain ($J')$ coupling regimes. In all cases, the 
staggered (normal) array always displays a massless 
(massive) behavior at $\langle M \rangle =1/2$. Finally, it is worth 
remarking that this study of the simplest  
$N=2$ leg ladder and with dimerization, ($p=2$ in the notation 
of \cite{CG}), can be  extended using the same bosonization 
methods to the study of $N$ leg ladders with arbitrary 
polymerization $p$ \cite{CG2}.

In summary, by using bosonization and numerical techniques, we have 
studied two interesting phenomena occurring in a dimerized two-leg ladder: 
(i) the closing of the gap at zero magnetization by means of a fine-tuning 
mechanism and, (ii) the disappearance of the $1/2$ plateau (gap at finite 
$\langle M \rangle$) by alternating the dimerization of the chains along 
the rung direction. As for the first issue, we presented a thorough
quantitative study of this phenomenon for AF systems, whereas
(ii) was studied for the first time here. 
On this latter respect, though $\rm NH_4CuCl_3$  under high magnetic fields
is supposed to be described  by a two-leg ladder system, its
magnetization curves does not show up neither 
$\langle M \rangle = 0$ nor $\langle M \rangle = 1/2$ 
plateaus \cite{Shiramura}. We trust the present paper will help
to convey a better understanding of these ladder materials.

It is a pleasure to acknowledge fruitful discussions with 
A.\ Dobry, A.\ Honecker, A.A\ Nersesyan, P.\ Pujol and V.\ Rittenberg.
The authors acknowledge partial financial support of 
CONICET, Fundaci\'on Antorchas and Deutsche Ausgleichbank.
D.C.C.\ would like to thank the DAAD for financial support under
the Visiting Professors Programme and the Physikalisches Institut der
Universit\"at Bonn for hospitality.
This work is done under ANPCyT project No.\ 03-00000-02249.

{\small  $\#$ On leave of absence from the Universidad Nacional de La Plata 
and Universidad Nacional de Lomas de Zamora, Argentina.}



\vspace{-1.5cm}
\begin{figure}
\hbox{%
\epsfxsize=4.1in
\epsffile{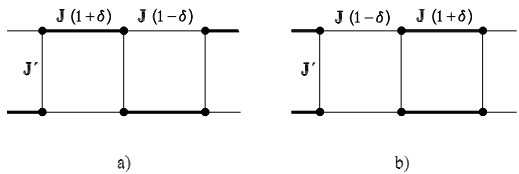}}
\vspace{-1.5cm}
\caption{Schematic view of alternating ladders with interchain coupling 
$J'$ and dimerization parameter $\delta$ for (a) staggered and, 
 (b) non-staggered arrays}
\end{figure}

\begin{figure}
\hbox{%
\hspace{-1.2cm}
\epsfxsize=3.5in
\epsffile{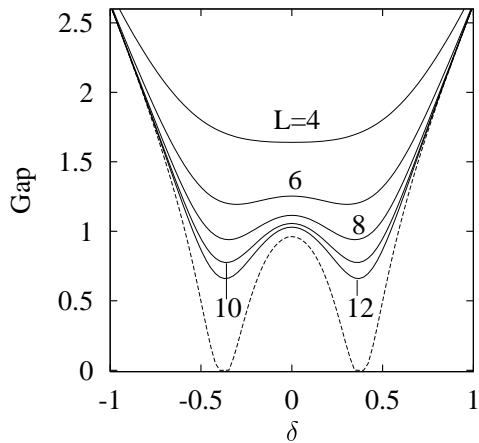}}
\vspace{-3cm}
\caption{Gap spectrum for different sizes of the staggered spin 
ladder for $J'/J = 1$. The dashed lines display extrapolations 
to the thermodynamic limit.}
\end{figure}

\vspace{-1cm}
\begin{figure}
\hbox{%
\epsfxsize=3.5in
\hspace{-1.2cm}
\epsffile{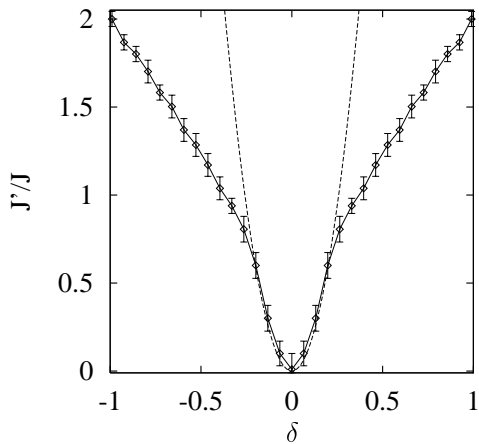}}
\vspace{-3.5cm}
\caption{Critical line of the staggered spin ladder in the
$(\delta ,J')$ coupling parameter space, extrapolated from finite samples.
Solid lines are guide to the eye whereas the dashed parabola stands
for the bosonization approach.}
\end{figure}

\vspace{9cm}
\begin{figure}
\hbox{%
\epsfxsize=3.5in
\hspace{-1.5cm}
\epsffile{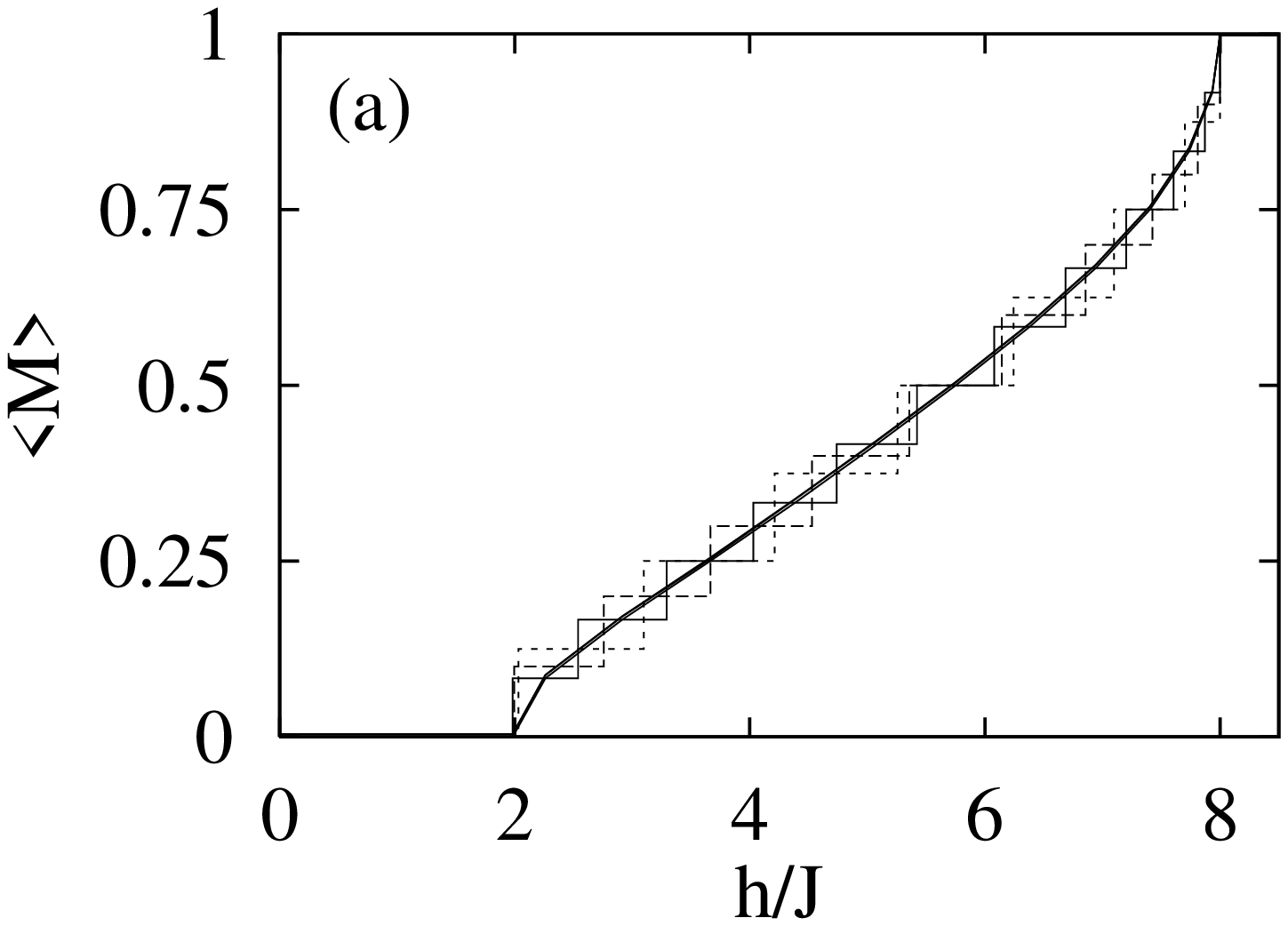}}
\end{figure}
\vspace{-1cm}
\begin{figure}
\hbox{%
\epsfxsize=3.5in
\hspace{-1.5cm}
\epsffile{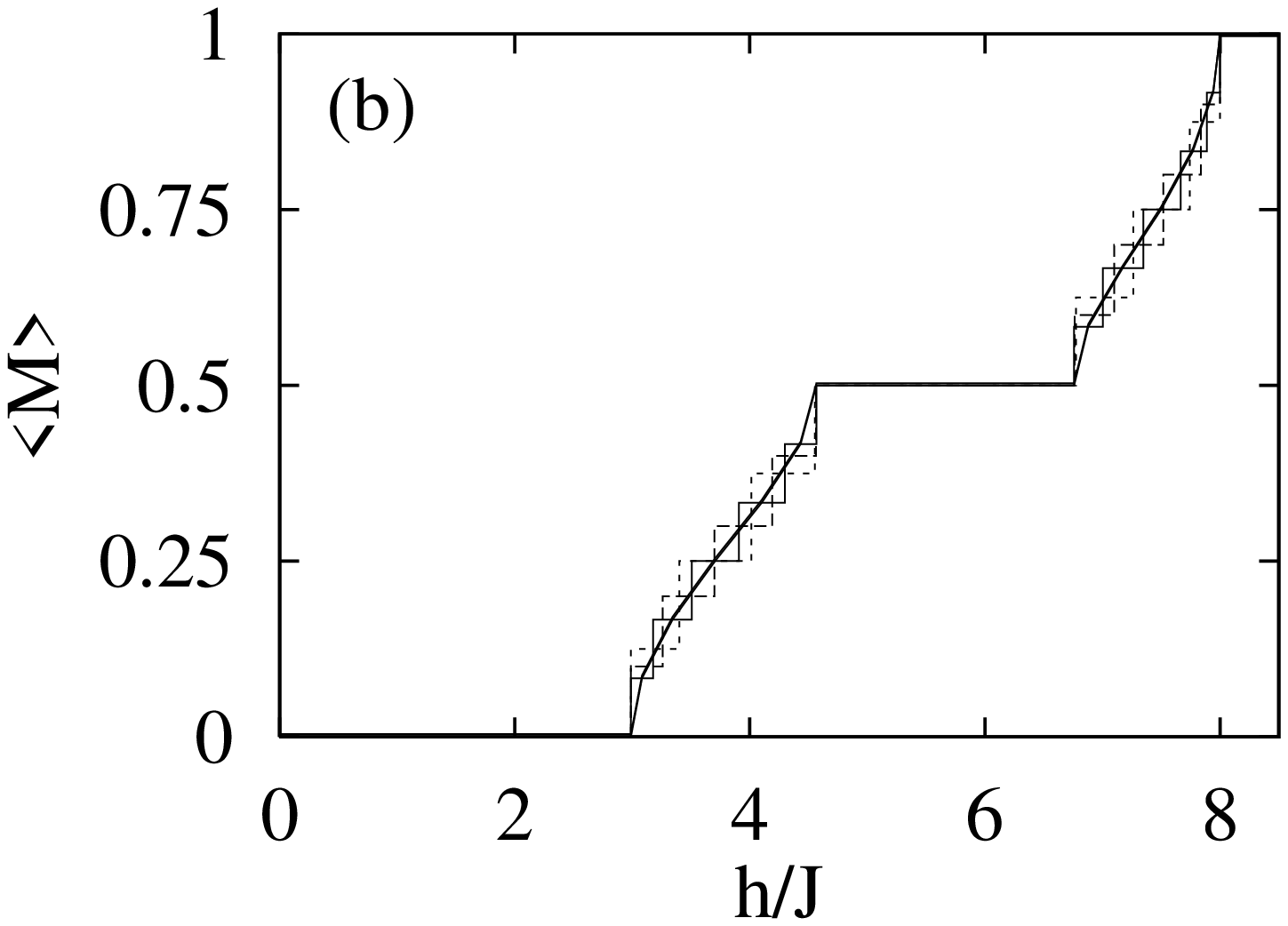}}
\caption{Magnetization curves of dimerized ladders for $\delta = 0.5$ and
$J'/J =2$ using both (a) staggered  and, (b) non-staggered 
coupling arrays. Solid, dashed and short dashed lines denote 
respectively magnetizations for $L=$ 12,10 and 8. 
The thick full line denotes the expected form in the 
thermodynamic limit}
\end{figure}


\begin{thebibliography}{99}

\bibitem{Elbio} The status of the field has been reviewed by T.M.\ Rice,
    Z.\ Phys. {\bf B 103}, 165 (1997); 
    E.\ Dagotto, T.M.\ Rice, Science {\bf 271}, 618 (1996).

\bibitem{CHP} D.C.\ Cabra, A.\ Honecker, P.\ Pujol,
                   Phys.\ Rev.\ Lett.\ {\bf 79}, 5126 (1997);
                Phys.\ Rev.\ {\bf B58}, 6241 (1998) and references therein.

\bibitem{OYA} M.\ Oshikawa, M.\ Yamanaka, I.\ Affleck,
                 Phys.\ Rev.\ Lett.\ {\bf 78}, 1984 (1997).   

\bibitem{Totsuka} K.\ Totsuka, Phys.\ Lett.\ {\bf A228}, 103 (1997).

\bibitem{Totsuka2} K.\ Totsuka, Phys.\ Rev.\ {\bf B57}, 3454 (1998).

\bibitem{CG} D.C.\ Cabra, M.D.\ Grynberg, Phys.\ Rev.\ {\bf B59}, 119 (1999).

\bibitem{Andreas} A.\ Honecker, Phys.\ Rev.\ {\bf B}
             in press (1999).

\bibitem{Shiramura} W.\ Shiramura  {\it et al.}, J.\ Phys.\ Soc.\ Jpn.\ {\bf
67}, 1548 (1998).

\bibitem{Sierra1} M.A.\ Mart\'\i n-Delgado, R.\ Shankar, G.\ Sierra,
          Phys.\ Rev.\ Lett.\ {\bf 77}, 3443 (1996).

\bibitem{Boucher} J. P. Boucher, L.P.\ Regnault, 
                  J. Phys. I (Paris) {\bf 6}, 1939 (1996).

\bibitem{Cross} M.C.\ Cross, D.S.\ Fisher, 
             Phys.\ Rev.\ {\bf  B19}, 402 (1979). See also
            J.\ Riera, A.\ Dobry Phys.\ Rev.\
             {\bf B51}, 16098 (1995); 
             G.\ Castilla, S.\ Chakravorty, V.J.\ Emery, 
             Phys.\ Rev.\ Lett.\ {\bf 75}, 1823 (1995).

\bibitem{DRS} E.\ Dagotto, J.\ Riera, D.J.\ Scalapino, Phys.\ Rev.\ 
             {\bf B47}, 5744 (1992).

\bibitem{SNT} K.\ Totsuka, M.\ Suzuki, J.\ Phys.\ Cond.\ Matt.\ {\bf 7}, 
6079 (1995); D.G.\ Shelton, A.A.\ Nersesyan, A.M.\ Tsvelik, 
Phys.\ Rev.\ {\bf B53}, 8521 (1996).


\bibitem{LeH}  I.\ Affleck,
          in {\it Fields, Strings and Critical Phenomena, Les Houches,
          Session XLIX}, edited by E.\ Brezin and J.\ Zinn-Justin
          (North-Holland, Amsterdam, 1988).

\bibitem{SchBo} H.J.\ Schulz, Phys.\ Rev.\ {\bf B34}, 6372 (1986).  
                         
\bibitem{HaldXXZ} F.D.M.\ Haldane, Phys.\ Rev.\ Lett.\ {\bf 45}, 1358
(1980).

\bibitem{Woyna} F.\ Woynarovich, H.-P.\ Eckle, T.T.\ Truong,
              J.\ Phys.\ A: Math.\ Gen.\ {\bf 22}, 4027 (1989).
\bibitem{Bogo} N.M.\ Bogoliubov, A.G.\ Izergin, V.E.\ Korepin,
          Nucl.\ Phys.\ {\bf B275}, 687 (1986).
\bibitem{bookK} V.E.\ Korepin, N.M.\ Bogoliubov, A.G.\ Izergin, {\it
Quantum Inverse Scattering Method and Correlation Functions}, Cambridge
          University Press, Cambridge (1993).

\bibitem{Lanczos} G.H.\  Golub, C.F.\ 
         Van Loan,{\it "Matrix Computations"}, 
          3rd ed. (Johns Hopkins University Press,  Baltimore 1996).


\bibitem{Gutt} A.J.\ Guttmann in 
            {\it Phase Transitions and Critical
            Phenomena}, edited by C.\ Domb, J.\ Lebowitz (Academic
            Press, New York  1990), Vol. 13. Consult also C.J.\ Hamer, 
            M.N.\ Barber, J.\ Phys.\ {\bf A14}, 2009 (1981).


\bibitem{Cardy} J.\ L.\ Cardy, J.\ Phys.\ {\bf A 17}, L385 (1984);
                  H.\ W.\ Bl\"ote, J.\ L.\ Cardy, M.\ P.\ Nightingale, 
                  Phys.\ Rev.\ Lett.\ {\bf 56}, 742 (1986) ; I.\ Affleck, 
Phys.\ Rev.\ Lett.\ {\bf 56}, 746 (1986).

\bibitem{CG2} D.C.\  Cabra, M.D.\ Grynberg, in preparation. 

\end{thebibliography}
\end{document}